\documentclass[aps,floatfix,nofootinbib]{revtex4}

\usepackage{amsmath,bm,amssymb,amsfonts,graphicx,psfrag}
\usepackage{ulem,color,xcolor}
\usepackage{epstopdf}

\allowdisplaybreaks

\begin{document}
\title{Tumbling of a Brownian particle in an extensional flow}

\author{Emmanuel Lance Christopher VI Medillo Plan}
\affiliation{Laboratoire Jean Alexandre Dieudonn\'e, Universit\'e Nice Sophia Antipolis, CNRS, UMR 7351,
06108 Nice, France}
\author{Dario Vincenzi}
\affiliation{Laboratoire Jean Alexandre Dieudonn\'e, Universit\'e Nice Sophia Antipolis, CNRS, UMR 7351,
06108 Nice, France}


\begin{abstract}
The phenomenon of tumbling of microscopic objects is commonly associated with shear flows. We address the question of whether tumbling can also occur in stretching-dominated flows. To answer this, we study the dynamics of a semi-flexible trumbbell in a planar extensional velocity field. We show that the trumbbell undergoes a random tumbling-through-folding motion. The probability distribution of long tumbling times is exponential with a time scale exponentially increasing with the Weissenberg number.
\end{abstract}

\maketitle

\section{Introduction}
Various microscopic objects, when immersed in a laminar shear flow, perform a tumbling motion in the plane of the shear. These include anisotropic solid particles \cite{EMLAAHM16}, flexible and semi-flexible polymers \cite{STSC05,Gerashchenko06,HWPKB13}, vesicles \cite{KS06}, bacteria \cite{KK09}, red blood cells \cite{DSV12} (for related numerical studies, see, e.g., Refs.~\cite{BM03,BKM05,YS07,HSGW11,WSB11,SG15}). The tumbling dynamics depends on the nature of the object and on its interaction with the fluid: tumbling may occur only within a restricted range of shear rates, and it can be periodic, chaotic, or random.

Axisymmetric solid particles and elastic dumbbells are among the simplest objects that tumble in shear flows. In a viscous simple shear flow of a Newtonian fluid, a neutrally buoyant axisymmetric particle  spends most of the time aligned with the direction of the flow and periodically reverses its orientation \cite{J22,B62}. If the axisymmetry of the particle is broken, the tumbling motion may become doubly-periodic or even chaotic \cite{HL79,YGR97,EMLAAHM16}. An elastic dumbbell, which consists of two beads joined by a spring, also performs an end-over-end tumbling motion in a shear flow \cite{Celani05,Chertkov05,Puliafito05,Turitsyn07}. Under the effect of Brownian fluctuations, the reversals occur at random times and are characterised by a transition from the stretched to the coiled state.  The distribution of the time intervals separating two reversals has an exponential tail with a time scale that decreases as a power-law of the Weissenberg number (the product of the amplitude of the velocity gradient and the relaxation time of the spring).

Here we address the question of whether tumbling can also exist in stretching-dominated flows and, if so, what the minimal requirements are for a particle to tumble in such flows and how the resulting tumbling motion compares with the analogous motion in a shear flow. The orientational dynamics of an axisymmetric particle or of an elastic dumbbell in a purely extensional velocity field is trivial: such objects indeed simply align with the stretching direction. In order to observe tumbling in an extensional flow, we must consider objects that can bend.
The `trumbbell' is one of the simplest semi-flexible objects:
It consists of three beads joined by two rigid connectors and of an elastic hinge at the central bead \cite{Hassager74} (see also \cite{Bird87}). Owing to its simplicity,
the trumbbell was originally used to study the low-frequency dynamics of stiff macromolecules analytically (e.g. \cite{R87,GdlT94}). 
We show that in an extensional flow a trumbbell spends a significant amount of time extended and oriented 
along the stretching direction of the flow; occasionally, a favourable sequence of Brownian fluctuations 
make the trumbbell fold, reverse its orientation, and unfold.
We examine this tumbling-through-folding dynamics
in terms of the stable configurations of the trumbbell and of the associated basins of attraction. 
The properties of the statistics of the tumbling times are explained by using the large deviations theory.
In particular, the analysis of the tumbling statistics reveals a fundamental difference between 
the tumbling motion of a trumbbell in an 
extensional flow and the tumbling motion of an elastic dumbbell in a simple
shear flow. In the former case, indeed, the mean time that separates two reversals grows with the Weissenberg 
number, and the growth is exponential.

Section \ref{model} describes the trumbell and specifies its configuration in terms of angular variables. Section \ref{statistics} examines the stationary statistics of the configuration of a trumbbell in a two-dimensional extensional flow. Section \ref{tumbling} analyses the tumbling motion and its statistical properties; analogous results in three dimensions are also presented. Finally, conclusions  are drawn in \S~\ref{conclusion}.

\section{The trumbbell}
\label{model}

\begin{figure}
  \centering
  \includegraphics[scale=0.95]{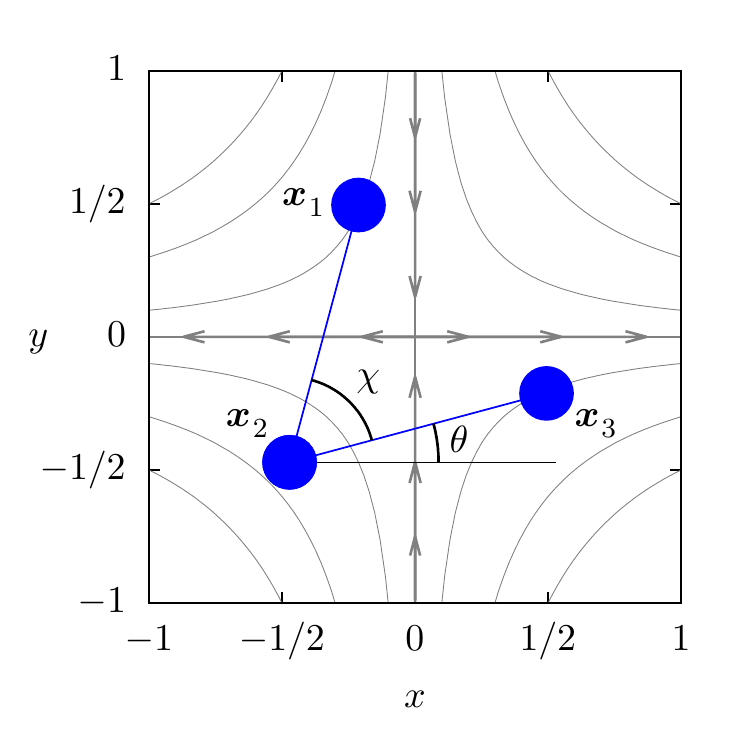}
  \caption{A trumbbell in the frame of reference of its centre of mass. The origin coincides with the centre of mass. The flow is extensional (streamlines shown in gray) with stretching in the $x$ direction and compression in the $y$ direction. (Online version in colour.)}
\label{fig:trumbbell}
\end{figure}

A trumbbell consists of three identical beads located at ${\bm x}_{\nu} \, (\nu=1,2,3)$ and joined by two inertialess rods of length $\ell$ (see figure \ref{fig:trumbbell})\footnote{Throughout the paper vectors are written in bold; calligraphic letters denote tensors.}. 
The internal angle between the rods is denoted by $\chi;$ since we distinguish the two configurations obtained by interchanging ${\bm x}_1$ and ${\bm x}_3$, $\chi$ varies between $0$ and $2\pi.$ When $\chi=0,2\pi$ the trumbbell is folded; when $\chi=\pi$ it is fully extended. A restoring force is imposed so that the trumbbell unfolds towards an equilibrium at $\chi=\pi.$ This force is described by the harmonic potential $\phi(\chi)=\mu(\chi-\pi)^2/2$ and models the resistance of the trumbell to bend. 

The trumbbell is advected by a linear velocity field ${\bm u}({\bm x},t),$ in which each bead experiences Stokes' drag  with drag coefficient $\zeta.$ The fluid is Newtonian. The beads are assumed to be sufficiently small to experience Brownian fluctuations. Furthermore, the inertia of the beads and the hydrodynamic interactions between them are neglected. It is also assumed that the trumbbell does not modify the flow.

The centre of mass of the trumbbell, ${\bm x}_{\mathrm{CM}}=({\bm x}_1+{\bm x}_2+{\bm x}_3)/3$, moves according to the equation:
$\dot{\bm x}_{\mathrm{CM}}=\bm u(\bm x_{\mathrm{CM}},t)+\sqrt{2\zeta k_B T/3}\,\hat{\bm \eta}(t)$
, where $k_B$ is the Boltzmann constant, $T$ is temperature, and $\hat{\bm \eta}(t)$ is white noise (see the appendix).
In the reference frame of ${\bm x}_{\mathrm{CM}}$, the configuration of the trumbbell can be described by $2(d-1)$ angular coordinates $\bm q,$  where $d$ is the dimension of the flow \cite{Hassager74}. The statistics of the configuration of the trumbbell at a time $t$ is specified by the probability density function (p.d.f.) $\psi(\bm q;t),$ which is normalised as $\int \psi({\bm q};t)\,d{\bm q}=1$ and satisfies the following diffusion equation \cite{Hassager74} (see also Ref.~\cite{Bird87} for the derivation):
\begin{eqnarray}
\frac{\partial\psi}{\partial t}=-\dfrac{\partial}{\partial q^i}\left\{{\mathcal G}^{ij}\left[\left(\kappa^{kl}(t)r_\nu^{l}\frac{\partial r_\nu^k}{\partial q^j}-\frac{1}{\zeta}\dfrac{\partial\phi}{\partial q^j}\right)\psi-\frac{k_BT}{\zeta}\sqrt{h}\frac{\partial}{\partial q^j}\left(\frac{\psi}{\sqrt{h}}\right)\right]\right\},
\label{eq:diffusion}
\end{eqnarray}
where $\kappa^{kl}(t)=\partial^l u^k(t)$ is the velocity gradient evaluated at $\bm x_{\mathrm{CM}},$ ${\bm r}_{\nu}={\bm x}_{\nu}-{\bm x}_{\mathrm{CM}}$ are the separation vectors describing the location of the beads with respect to $\bm x_{\mathrm{CM}},$ $h=\mathrm{det}({{\mathcal H}})$ with ${\mathcal H}^{ij}=\sum_{\nu,k} \frac{\partial r^k_\nu}{\partial q^i} \frac{\partial r^k_\nu}{\partial q^j},$ and ${{\mathcal G}}={{\mathcal H}}^{-1}.$ Summation over repeated indices is implied in \eqref{eq:diffusion}.
Note that in a linear flow the orientational dynamics of the trumbbell only depends on the velocity gradient, which is spatially uniform (see also the evolution equations for the configuration of the trumbbell in the appendix). 
The internal configuration of the trumbbell is therefore decoupled from the
position of the centre of mass. This implies that the statistics of $\bm q$ is independent of $\bm x_{\mathrm{CM}}$;
hence, without any loss of generality,
we study the dynamics of the trumbbell in the frame of reference that is translated with the centre of mass.

To simplify the analysis, we assume that the trumbbell is immersed in a thin layer of fluid and we restrict our attention to $d=2$. The case of a planar three-dimensional flow will be discussed at the end of \S~\ref{tumbling}. In two dimensions, we choose ${\bm q}=(\theta,\chi),$ where $0\leq\theta< 2\pi$ is the angle that the vector ${\bm x}_3-{\bm x}_2$ makes with the $x$-axis; $\theta$ gives the orientation of the trumbbell in the plane. The vectors $\bm r_{\nu}$ can then be expressed in terms of $\theta$ and $\chi$ as
\begin{equation}
\begin{array}{l}
\bm r_1=\ell\left(2\cos(\theta+\chi)-\cos(\theta),
2\sin(\theta+\chi)-\sin(\theta)\right)/3,\\
\bm r_2=-\ell\left(\cos(\theta+\chi)+\cos(\theta),
\sin(\theta+\chi)+\sin(\theta)\right)/3,\\
\bm r_3=\ell\left(2\cos(\theta)-\cos(\theta+\chi),
2\sin(\theta)-\sin(\theta+\chi)\right)/3.
\end{array} \nonumber
\end{equation}
We also have $h=\ell^4(4-\cos^2\chi)/9$ and   
\begin{eqnarray}
\mathcal{G}=\ell^{-2}\begin{pmatrix}
\dfrac{6}{4-\cos^2\chi} & -\dfrac{3}{2+\cos\chi}
\\[4mm]
-\dfrac{3}{2+\cos\chi} & \dfrac{6}{2+\cos\chi}
\end{pmatrix}. \nonumber
\end{eqnarray}
Equation (\ref{eq:diffusion}) can then be rewritten as a Fokker--Planck equation in two variables:
\begin{eqnarray}
\partial_t \psi=-\partial_\theta (V_\theta\psi)-\partial_\chi (V_\chi\psi)
+\textstyle\frac{1}{2}\partial_\theta^2(D_{\theta\theta}\psi)+\partial_\theta
\partial_\chi(D_{\theta\chi}\psi)
+\textstyle\frac{1}{2}\partial_\chi^2(D_{\chi\chi}\psi),
\label{eq:FPE-2D}
\end{eqnarray} where
\begin{equation}
D_{\theta\theta}=\dfrac{12 k_BT}{\zeta\ell^2(4-\cos^2\chi)},\qquad
D_{\theta\chi}=-\dfrac{6 k_BT}{\zeta\ell^2(2+\cos\chi)},\qquad
D_{\chi\chi}=\dfrac{12 k_BT}{\zeta\ell^2(2+\cos\chi)} \nonumber
\end{equation}
and
\begin{eqnarray}
\nonumber
V_{\theta}&=&-\dfrac{6k_BT\sin\chi}{\zeta\ell^2(2-\cos\chi)(2+\cos\chi)^2}-\dfrac{3\mu(\pi-\chi)}{\zeta\ell^2(2+\cos\chi)}-\\ \nonumber
&&\dfrac{1}{4-\cos^2\chi}\{[2\sin\chi\cos^2(\theta +\chi)+\cos\theta(4\sin\theta-\cos\chi\sin(\theta+\chi))]\kappa^{11}(t)+ \\ \nonumber
&&[2\cos\chi\sin^2(\theta+\chi)-\sin\theta(2+\cos\chi)\sin(\theta+\chi)+4\sin^2\theta]\kappa^{12}(t)+ \\ \nonumber
&&[-2\cos\chi\cos^2(\theta+\chi)+\cos\theta(2+\cos\chi)\cos(\theta+\chi)-4\cos^2\theta]\kappa^{21}(t)+ \\ \nonumber &&[2\cos\theta(\sin(\theta+\chi)-2\sin\theta)+\cos\chi(\sin\theta\cos(\theta+\chi)-\sin(2(\theta+\chi)))]\kappa^{22}(t)\},
\\[2mm] \nonumber
V_{\chi}&=& \dfrac{12k_BT\sin\chi}{\zeta\ell^2(2-\cos\chi)(2+\cos\chi)^2}+ \dfrac{6\mu(\pi-\chi)}{\zeta\ell^2(2+\cos\chi)}+ \dfrac{\sin\chi} {2+\cos\chi}\left\{[1-2\cos(2\theta+\chi)]\kappa^{11}(t)- \right. \\ \nonumber
&& \left. 2\sin(2\theta+\chi)[\kappa^{12}(t)+\kappa^{21}(t)]+ [1+2\cos(2\theta+\chi)]\kappa^{22}(t)\right\}.
\end{eqnarray}
The Fokker--Planck equation (\ref{eq:FPE-2D}) is equivalent to the following system of It\^{o} stochastic differential equations:
\begin{equation}
\begin{split}
\dot{\theta}(t)&=V_\theta +\sqrt{D_{\theta\theta}}\,\xi_\theta(t), \\
\dot{\chi}(t)&=V_\chi +\dfrac{D_{\theta\chi}}{\sqrt{D_{\theta\theta}}}\,\xi_\theta(t)+\sqrt{D_{\chi\chi}-\dfrac{D^2_{\theta\chi}}{D_{\theta\theta}}}\,\xi_\chi(t),
\label{eq:dthetachi}
\end{split}
\end{equation}
where $\xi_\theta(t)$ and $\xi_\chi(t)$ are independent white noises. Note that in \eqref{eq:dthetachi} $D_{\chi\chi}-{D^2_{\theta\chi}}/{D_{\theta\theta}}=3k_BT/\zeta \ell^2.$

\section{Stationary statistics of the configuration}
\label{statistics}

By rescaling time in system (\ref{eq:dthetachi}), it is possible to formulate the equations in terms of two dimensionless parameters: the stiffness parameter $Z=\mu/k_BT$ and the Weissenberg number $\mathit{Wi}=\gamma \zeta \ell^2/\mu,$  where $\gamma$ is the magnitude of the velocity gradient. The former expresses the relative intensity of the restoring force to that of Brownian noise, whereas the latter compares the strengths of the flow and of the restoring force. In addition, we introduce the P\'{e}clet number $\mathit{Pe}=\gamma \zeta \ell^2/k_BT=Z\mathit{Wi},$ which is the relative intensity of the flow to Brownian noise.

\begin{figure}
\centerline{
\includegraphics[width=0.53\textwidth]{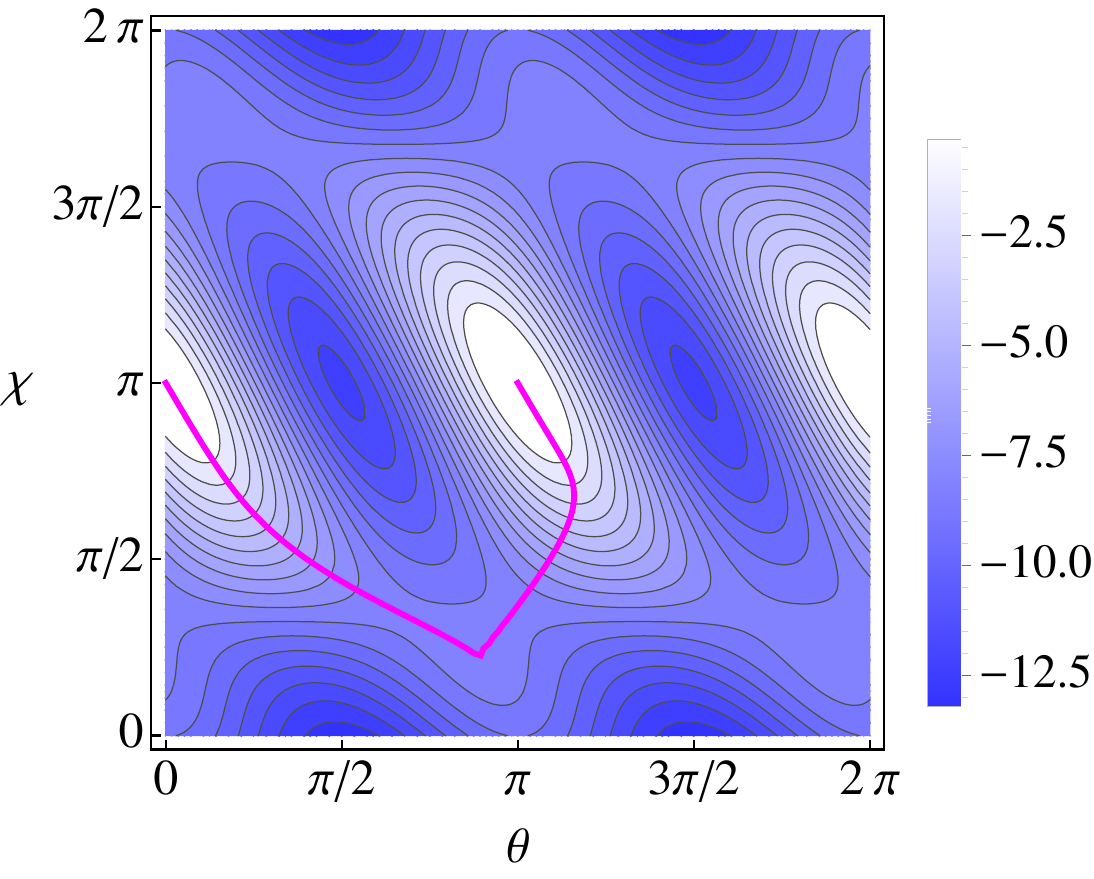}
\includegraphics[width=0.47\textwidth]{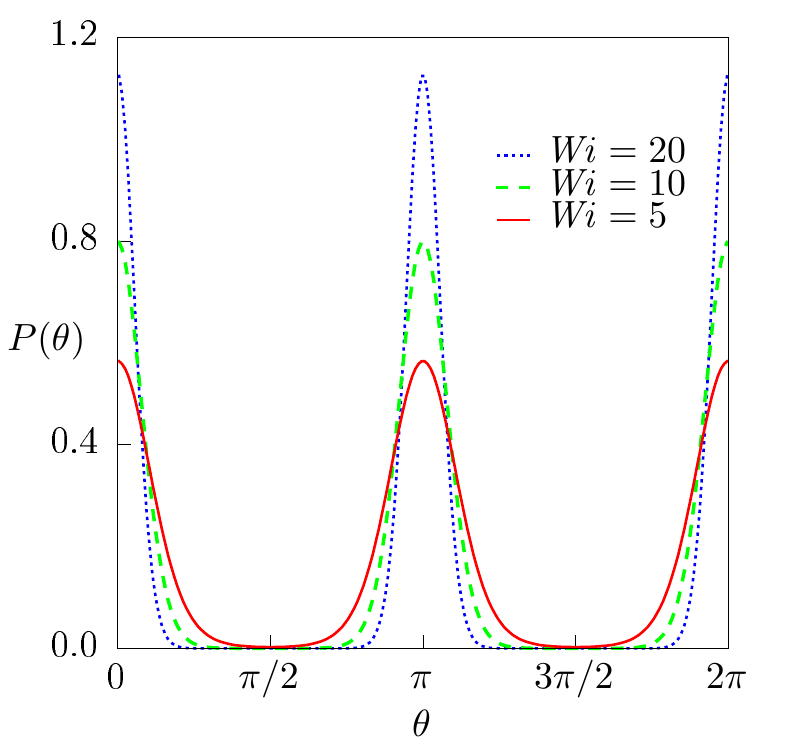}
}
\caption{Left: Contourplot of $\ln \psi_{\textrm{st}}(\theta,\chi)$ for $Z=1$ and $\mathit{Wi}=6.$ One minimum energy path (see \S~\ref{tumbling}) connecting $(\pi,\pi)$ and $(0,\pi)$ is drawn with a solid magenta curve. Right: The marginal p.d.f. of $\theta$ for $Z=1$ and $\mathit{Wi}=5$ (red, solid), $\mathit{Wi}=10$ (green, dashed), and $\mathit{Wi}=20$ (blue, dotted). (Online version in colour.)}
\label{fig:pdf}
\end{figure}

Consider a two-dimensional extensional flow ${\bm u}(x,y)=\gamma(x,-y), \gamma>0,$ which consists of a stretching direction $x$ and a compressing direction $y$ (figure \ref{fig:trumbbell})\footnote{As noted in \S~2, our study is not restricted to the case in which the centre of mass stays fixed at the stagnation
point of the extensional flow, but it is conducted in the frame of reference of the centre of mass and takes advantage of
the fact that the statistics of the configuration is independent of $\bm x_{\mathrm{CM}}$.}. For this flow, the stationary p.d.f. of $\theta$ and $\chi$ takes the form $\psi_{\textrm{st}}(\theta,\chi)=N\sqrt{h}\exp{\left[(\Phi-\phi)/(k_BT)\right]}$ with $\Phi=(\zeta/2)\sum{\kappa^{ij}r_{\nu}^ir_{\nu}^j}$ \cite{Hassager74}; its explicit expression is:
\begin{equation}
\psi_{\textrm{st}}(\theta,\chi)=N\sqrt{4-\cos^2 \chi}\exp\left[\frac{Z\,\mathit{Wi}(2\cos\chi-1)\cos(2\theta+\chi)}{3}-\frac{Z(\pi-\chi)^2}{2}\right],
\label{eq:pdf}
\end{equation}
where $N$ is a normalization constant. The contour plot of $\ln \psi_{\textrm{st}}(\theta,\chi)$ is shown in figure \ref{fig:pdf} (left) for representative values of $Z$ and $\mathit{Wi}.$ 
It is easily shown that the maxima of $\psi_{\textrm{st}}(\theta,\chi)$ are located at $P_{\pi}:(\theta=n\pi, \chi=\pi),$ where $n$ is an integer. Hence, the trumbbell spends most of the time in an extended configuration and oriented in the stretching direction of the flow. The peaks at $P_{\pi}$ become narrower as $\mathit{Wi}$ increases, showing stronger preference for these configurations (see the marginal p.d.f. of $\theta$, $P(\theta)$ in figure \ref{fig:pdf}, right). Also note that these peaks are not symmetrically distributed around $P_{\pi}.$ 

To understand the behaviour of $\psi_{st}(\theta,\chi)$ when the flow is much stronger than both the restoring force and Brownian noise, we set $\mu = 0$ and $k_BT=0$ in (\ref{eq:dthetachi}), which corresponds to considering the limiting case $\mathit{Wi}= \mathit{Pe}= \infty.$ The resulting system is: 
\begin{equation}
\begin{split}
\dot{\theta}(t)&=-\dfrac{\gamma}{2(4-\cos^2\chi)} \left[7 \sin(2\theta)+4\cos(2\theta+\chi)\sin \chi-\sin(2(\theta+\chi)) \right],\\
\dot{\chi}(t)&= -\dfrac{4 \gamma }{2+\cos \chi}\left[\sin \chi \cos(2\theta+\chi)\right].
\end{split}
\label{eq:stabanal}
\end{equation}
The linear stability analysis of this system yields two stable configurations, both aligned with the stretching direction of the flow $(\theta=0,\pi,2\pi).$ The first stable configuration is the extended one $P_{\pi}$ and is characterised by two negative eigenvalues: $\lambda_{\pi}^{(1)}=-4\gamma,\lambda_{\pi} ^{(2)}=-2\gamma.$ The second stable configuration is the folded one $P_0:(\theta=n\pi,\chi=0,2\pi)$ with eigenvalues $\lambda_{0}^{(1)}=-2\gamma,\lambda_{0}^{(2)}=-4\gamma/3.$ As the velocity gradient $\gamma$ becomes stronger, both configurations become increasingly stable, since the eigenvalues are proportional to $\gamma.$ However, the ratio of the most negative eigenvalues of the two configurations is $\lambda_{\pi}^{(1)}/\lambda_0^{(1)}=2;$ hence the extended configuration $P_{\pi}$ is more stable than the folded one $P_{0}$ for all $\gamma$ and is expected to dominate the long-time statistics. This fact can be understood by noting that the velocity of a bead is proportional to its distance from the center of mass, and in the extended configuration $P_{\pi}$ the end beads are farther from ${\bm x}_{\mathrm{CM}}$ than they are in the folded configuration $P_{0}$. 

\begin{figure}
\centering
\includegraphics[width=0.5\textwidth]{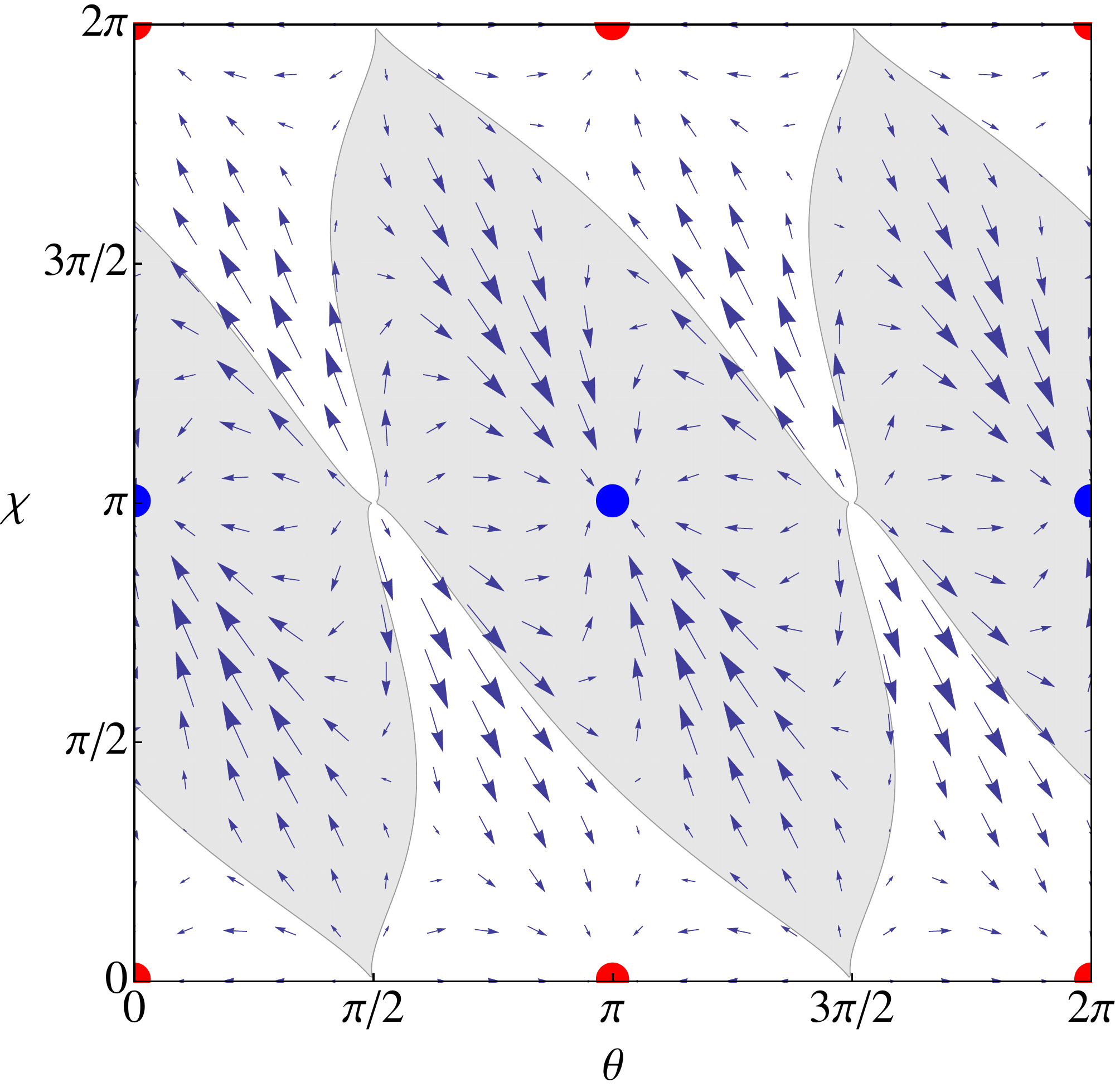}%
\hfill
\includegraphics[width=0.5\textwidth]{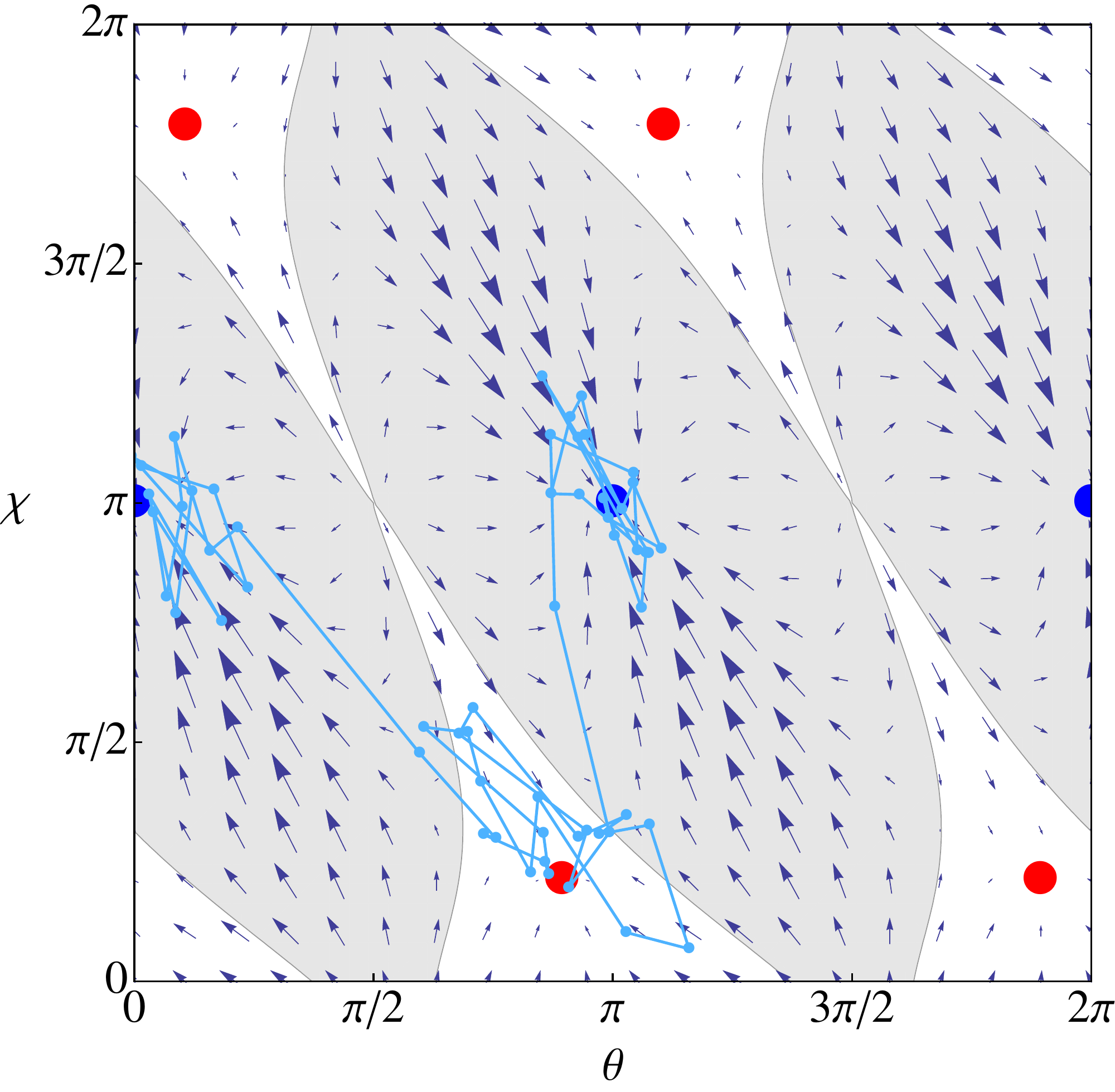}
  \caption{Left: Vector plot of $\gamma^{-1}(\dot{\theta},\dot{\chi})$ for $\mathit{Wi}=\mathit{Pe}=\infty.$ The size of the arrows are proportional to the magnitude of the vector $\gamma^{-1}(\dot{\theta},\dot{\chi}).$ The basins of attraction of $P_{\pi}$ (blue disks) are in gray; those of $P_{0}$ (red disks) are in white. Right: Vector plot of $(\dot{\theta},\dot{\chi})$ for $Z=1, \mathit{Wi}=6.$ The white areas are now the basins of attraction of $P_{\star}$ (red disks). The light blue curve with markers is a trajectory $(\theta(t),\chi(t))$ corresponding to a tumbling motion from $\theta=\chi=\pi$ to $\theta=0,\chi=\pi$. (Online version in colour.)}
    \label{fig:traj}
\end{figure}

\begin{figure}
  \centerline{\includegraphics[scale=0.7]{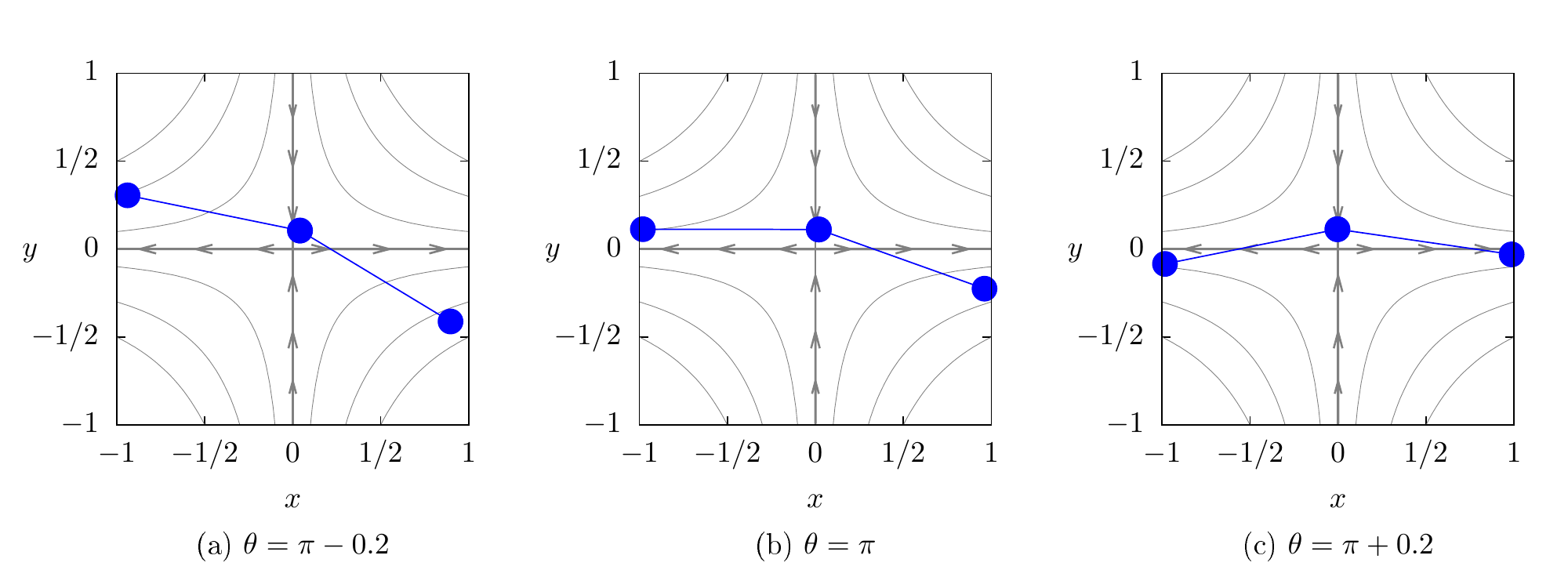}}
  \caption{Three configurations near an extended equilibrium configuration for the same value of $\chi=2.8.$ (Online version in colour.)}
\label{fig:pos}
\end{figure}

The presence of the two stable configurations is seen in the vector plot of $(\dot{\theta},\dot{\chi})$ shown in figure \ref{fig:traj} (left). 
As both $\dot{\theta}$ and $\dot{\chi}$ are proportional to $\gamma$ [see \eqref{eq:stabanal}], the geometrical structure of the vector plot does not change with $\gamma.$ In accordance with the stability analysis of the fixed points of (\ref{eq:stabanal}), the vectors that lie in the neighbourhood of $P_{\pi}$ are larger than those in the neighbourhood of $P_0.$ Therefore, the presence of Brownian noise allows an easier escape from the basin of attraction of the folded configuration $P_{0}$ than from that of the extended configuration $P_{\pi}$. Conversely, a trumbbell in an extended configuration is more likely to remain in the basin  of attraction of this configuration, until there is sufficient noise for it to fold. Also note that the basin of attraction of the points $P_{\pi}$ is not symmetrically distributed around them, but there is a preferential direction along which the system is more strongly attracted (see figure \ref{fig:traj}). Consider indeed two configurations with the same value of $\chi$ close to $\pi$ and with $\theta$ either slightly less than $\pi$ or slightly greater than it (see figure \ref{fig:pos}). It is clear that the latter configuration is more strongly attracted to $P_{\pi}.$
\begin{figure}
  \centering \includegraphics[width=\textwidth]{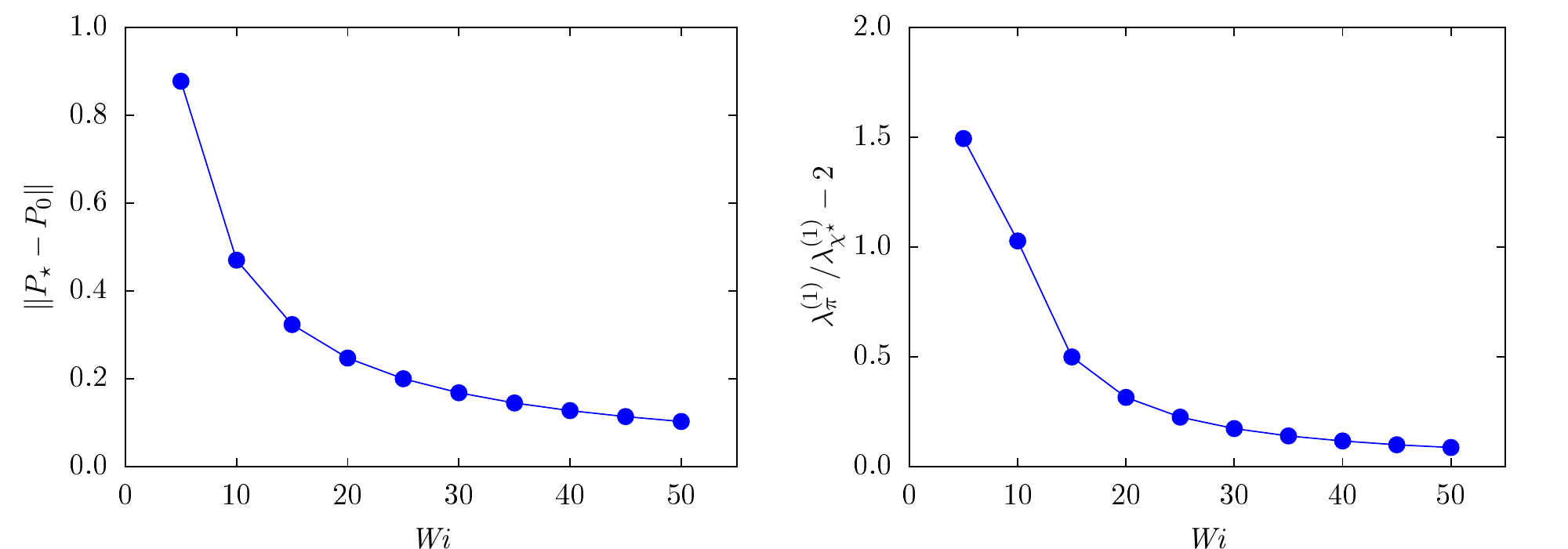}
  \caption{Left: Distance of a stable fixed point $P_{\star}$ from the nearest $P_0$ for $\mathit{Pe} = \infty$ as a function of $\mathit{Wi}.$ Right: Ratio of the most negative eigenvalue of $P_{\pi}$ to that of $P_{\star}$ subtracted by $\lambda_{\pi}^{(1)}/\lambda_0^{(1)}=2.$ (Online version in colour.)}
\label{fig:limit}
\end{figure}

In the presence of restoring potential $(\mathit{Wi}<\infty, \mathit{Pe} = \infty),$ the fixed points of system (\ref{eq:dthetachi}) and their stability can be calculated numerically. Only $P_{\pi}$ remain stable fixed points; the configurations $P_0$ are no longer fixed because of the restoring potential. However, if the flow is sufficiently strong ($\mathit{Wi} \gtrsim 4$ for $Z=1$), there exist stable points $P_{\star}:(\theta_{\star},\chi_{\star})$ that approach $P_0$ as $\mathit{Wi}$ increases (see figures \ref{fig:traj}, right and figure \ref{fig:limit}, left). Moreover, the points $P_{\pi}$ are more stable than $P_{\star}$ for all $\gamma$ and as $\mathit{Wi}$ increases, the eigenvalues of $P_{\pi}$ and $P_{\star}$ approach the corresponding eigenvalues of the $\mathit{Wi} = \infty$ case (figure \ref{fig:limit}, right). Finally, the basins of attraction of $P_{\pi}$ and $P_{\star}$ have similar structures to those of $P_{\pi}$ and $P_0$ when $\mathit{Wi}$ is infinite (figure \ref{fig:traj}, right). These results indicate that the intuition gained from the study of the $\mathit{Wi} = \infty$ case holds true also for $\mathit{Wi}<\infty.$ In particular, the stability analysis of the fixed points of the system explains why $\psi_{\textrm{st}}(\theta,\chi)$ shows high peaks only at the extended configuration $P_{\pi}$ and not at the folded configuration. Furthermore, the examination of the basins of attraction of $P_{\pi}$ and of the vector field $(\dot{\theta},\dot{\chi})$ clarifies the shape of these peaks.

\section{Tumbling dynamics}
\label{tumbling}

\begin{figure}
  \centerline{\includegraphics[width=1.03 \textwidth]{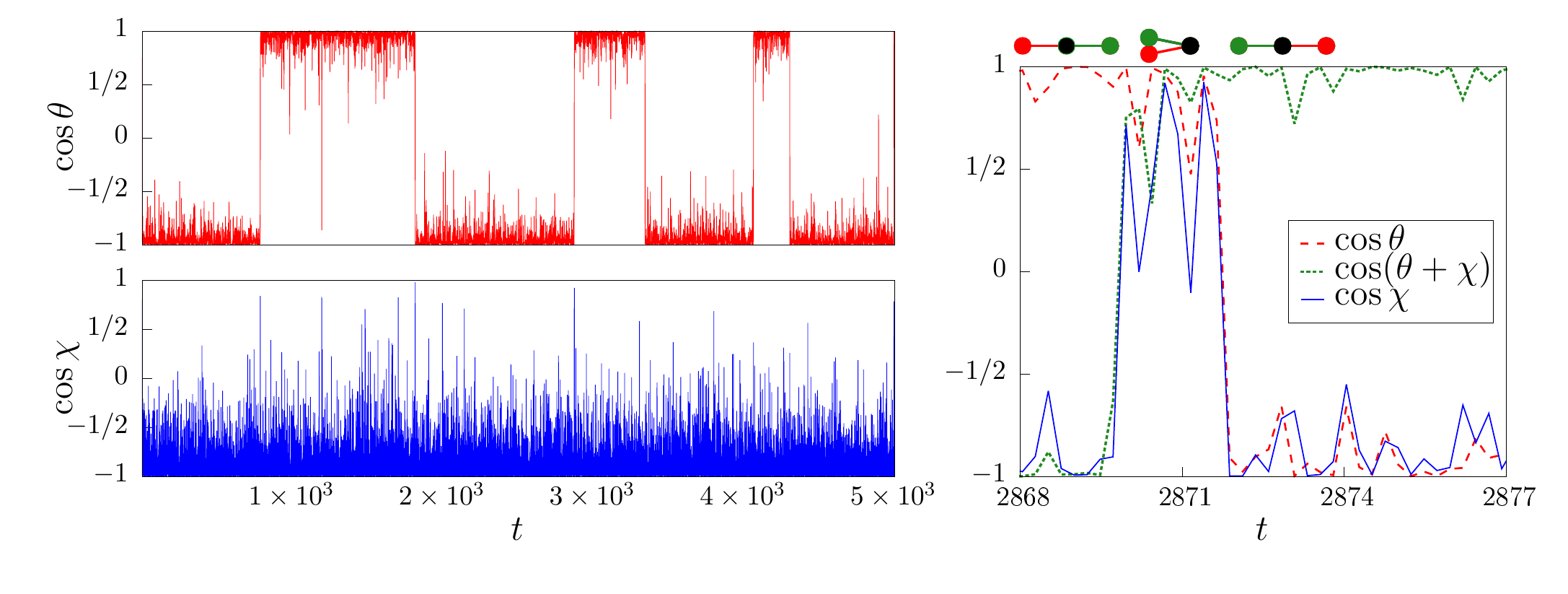}}
\caption{Time series of $\cos \theta(t)$ (left, top) and $\cos\chi(t)$ (left, bottom) for a two-dimensional extensional flow with $Z=1, \mathit{Wi}=6.$ Right: Zoom of the time series of $\cos \theta, \cos \chi, \cos (\theta+\chi)$ over a tumbling event. (Online version in colour.)}
\label{fig:tseries}
\end{figure}
For sufficiently large $\mathit{Wi}$, the trumbbell is trapped in the basin of attraction of one of the extended configuration $P_{\pi}$ for a long time. However, Brownian fluctuations may occasionally make it tumble between the aligned configuration ($\theta=0,2\pi$) and the anti-aligned one ($\theta=\pi$). To investigate this phenomenon, we have performed numerical simulations of (\ref{eq:dthetachi}) by using the Euler--Maruyama scheme with a time step $dt=10^{-3}$ \cite{KP99} (we also performed numerical simulations with smaller time steps, which confirmed the accuracy of our results). The inspection of a representative time series of $\cos \theta(t)$ and $\cos \chi(t)$ confirms the aforementioned tumbling dynamics and shows that these reversals are characterised by partial folding (figure \ref{fig:tseries}). A typical trajectory in the $(\theta,\chi)$ phase space is shown in figure \ref{fig:traj} (right). The trumbbell is initially extended and anti-aligned with the stretching direction. A favourable sequence of Brownian fluctuations makes it exit from the initial basin of attraction and pass through that of the folded configuration. The trumbbell then unfolds towards the extended but aligned configuration. The movie included as a supplementary material further illustrates this tumbling-through-folding motion.

Although the trumbbell always folds (possibly not completely) during a tumbling event, there are instances in which folding does not result into a reversal and the trumbbell rapidly unfolds back into the original configuration (see for instance figure \ref{fig:tseries} at $t \approx 1.2 \times 10^3$). In order to correctly identify a tumbling event, we shall therefore apply the following criterion. Suppose that at time $t_1$ the trumbbell is sufficiently extended, i.e. 
\begin{eqnarray}
|\cos \theta|>1-\epsilon, \quad |\cos (\theta+\chi)|>1-\epsilon, \quad \cos \theta \cos (\theta+\chi)<0
\label{eq:criteria}
\end{eqnarray}
for some small $\epsilon.$ We shall say that a tumbling occurs at time $t_2$ if $t_2$ is the smallest time after $t_1$ such that \eqref{eq:criteria} is again satisfied and $\cos \theta(t_1)\cos \theta(t_2)<0.$ In our simulations, we set $\epsilon=0.01$; we have verified that the specific choice of $\epsilon$ does not affect the statistical analysis of tumbling, provided the threshold $1-\epsilon$ is sufficiently close to 1.

\begin{figure}
  \centerline{\includegraphics[scale=0.82]{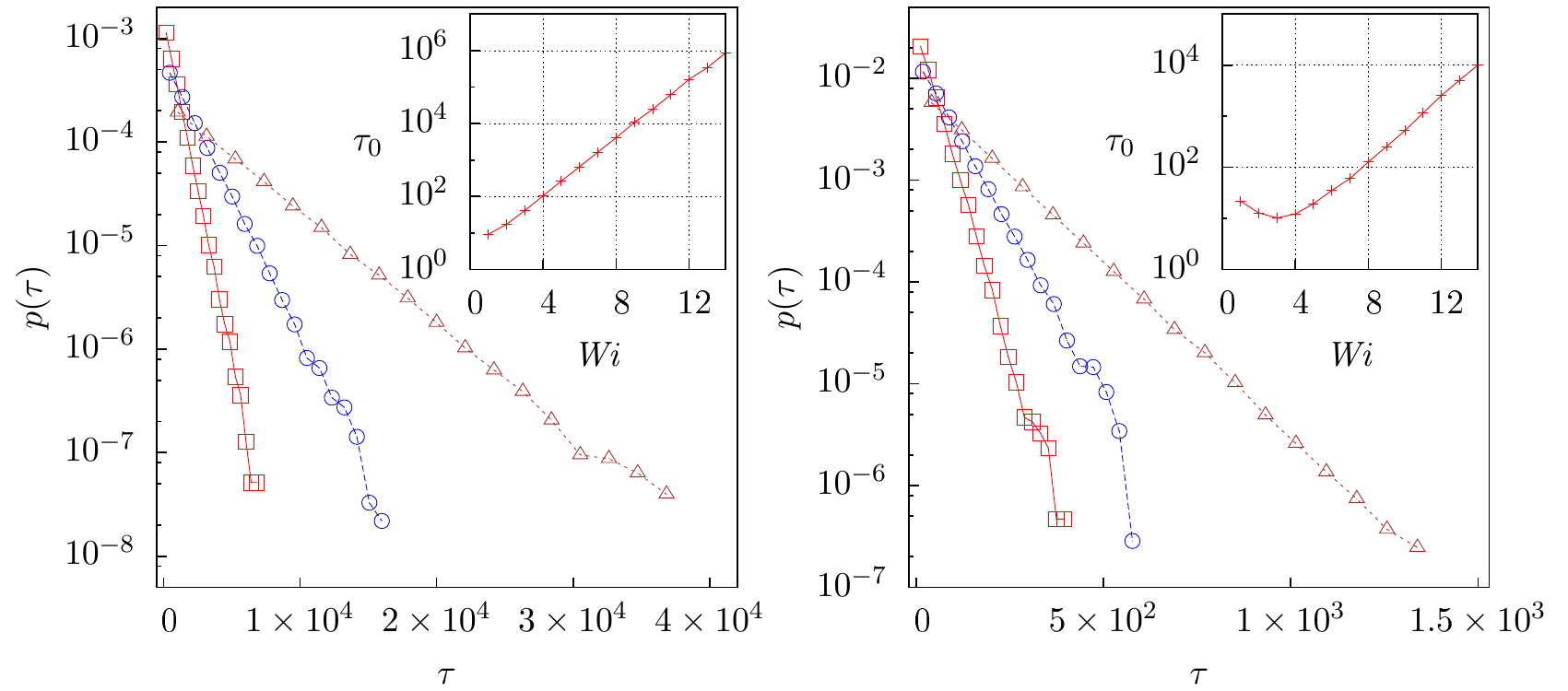}}
  \caption{P.d.f.s of the tumbling time in two dimensions (left) and three dimensions (right) for $Z=1$ and $\mathit{Wi}=6$ (red squares), $\mathit{Wi}=7$ (blue circles), and $\mathit{Wi}=8$ (brown triangles). The insets show $\tau_0$ as a function of $\mathit{Wi}$. (Online version in colour.)}
\label{fig:pdftumbling}
\end{figure}
The tumbling dynamics described above is not periodic. On the contrary, the time $\tau$ separating two tumbling events are distributed randomly. For each different value of $\mathit{Wi}$, we have collected at least $10^4$ tumbling realizations. We find that the p.d.f. of $\tau$ has an exponential tail for large $\tau: p(\tau)\propto \exp(-\tau/{\tau_0})$ for $\tau \gg \tau_0.$ Moreover, $\tau_0$ increases exponentially as a function of $\mathit{Wi}$ (figure \ref{fig:pdftumbling}, left). Hence, as the flow becomes stronger, it takes a longer time for a tumbling to occur. The configurations $P_{\pi}$ indeed become increasingly stable, and larger Brownian fluctuations are required for the system to escape from the basins of attraction of $P_{\pi}.$ 

The above properties of $p(\tau)$ can be predicted by using the Freidlin--Wentzell large deviations theory \cite{FW12} (see also \cite{T09}). Indeed, for large values of $\mathit{Wi}$, tumbling in an extensional flow can be regarded as escaping from an attractor of a stochastic dynamical system in the limit of small noise. The p.d.f. of the exit time thus has an exponential tail and the mean exit time increases exponentially as the amplitude of the noise vanishes. The same theory also predicts the tumbling-through-folding phenomenon. Indeed, the most probable transition paths, or minimum energy paths, that connect two adjacent configurations $P_{\pi}$ are parallel to the gradient of the pseudo-potential $V=-\ln \psi_{\mathrm{st}}$ \cite{EV10}. The application of the improved string method to $V$ \cite{ERV07} shows that these minimum energy paths pass through the folded configurations $P_{\star}$, which are saddle points of $V$. By way of illustration, a minimum energy path connecting $(\pi,\pi)$ and $(0,\pi)$ is given in figure \ref{fig:traj} (right).

The analysis so far considers a two-dimensional velocity field.
\begin{figure}
 \centerline{\includegraphics[width=1.03\textwidth]{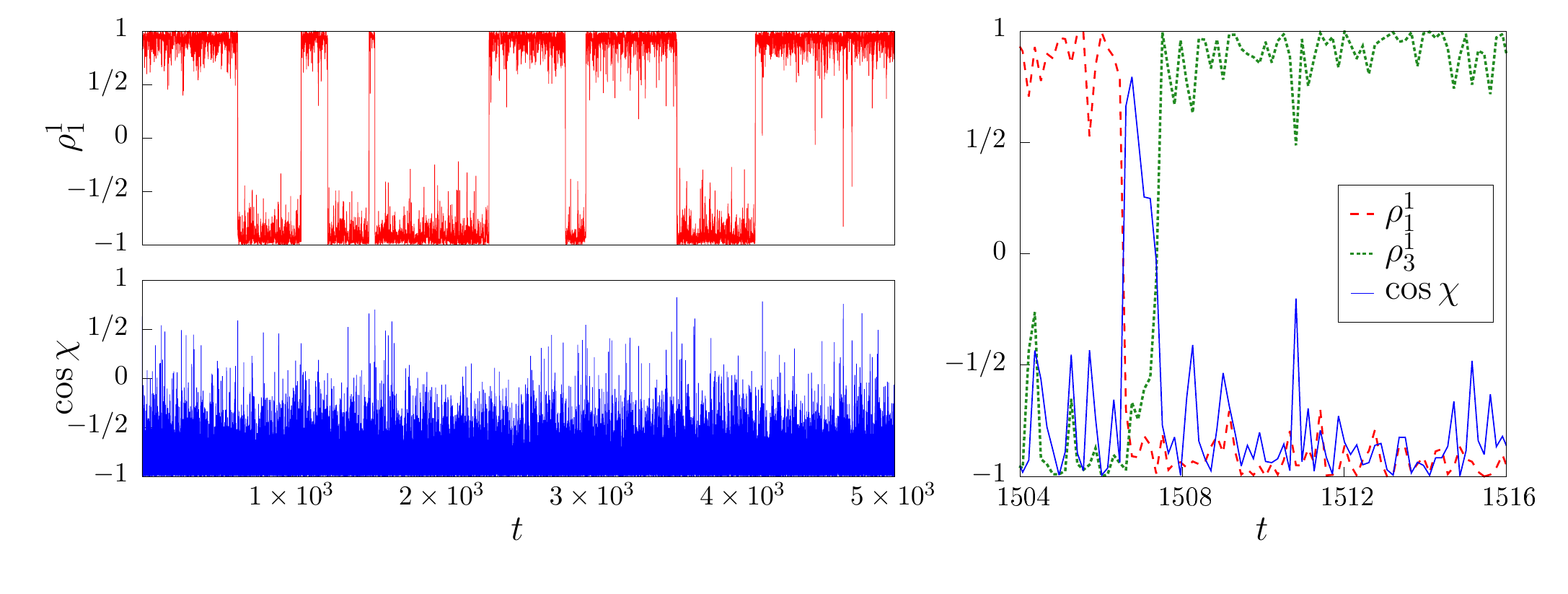}}
\caption{Time series of $\rho_1^1(t)$ (left, top) and $\cos\chi(t)$ (left, bottom) for a three-dimensional planar extensional flow with $Z=1, \mathit{Wi}=10.$ Right: Zoom of the time series of $\rho_1^1, \rho_3^1, \cos \chi$ over a tumbling event. (Online version in colour.)}
\label{fig:tseries3D}
\end{figure}
However, in a realistic planar extensional flow, a trumbbell may move outside the plane of the flow. This situation can be described by considering a three-dimensional velocity field ${\bm u}=\gamma(x,-y,0)$. In this case, the trumbbell has four degrees of freedom, which can be described by the internal angle $\chi$ and three Euler angles giving the orientation of the particle in space \cite{Hassager74,Bird87}. This formulation, however, is not suitable for numerical simulations because of singularities near to the extended configuration (see the form of the tensor $\mathcal G$ in\cite{Bird87}, p. 618). To circumvent this difficulty, we introduce the vectors ${\bm \rho}_1={\bm x}_1-{\bm x}_2$ and ${\bm \rho}_3={\bm x}_3-{\bm x}_2,$ which give the orientation of the rods in the frame of reference of the centre of mass. The system of Stratonovich stochastic differential equations that describes the time evolution of ${\bm \rho}_1$ and ${\bm \rho}_3$ is derived in the appendix. This system has been solved by using the Euler--Heun method with $dt=10^{-3}$ \cite{KP99}. When $\mathit{Wi}$ is sufficiently high, the trumbbell is observed to be predominantly in the extended configuration $\chi=\pi$ with the separation vectors aligned or anti-aligned with the stretching direction of the flow. At random times, the trumbbell folds and reverses its orientation, with a dynamics similar to that observed in two dimensions. A time series showing $\cos \chi$ and the first component of ${\bm \rho}_1$ is shown in figure \ref{fig:tseries3D}. The tumbling events can be identified as before by replacing $\cos \theta,\cos(\theta+\chi)$ with the first components of ${\bm \rho}_1,{\bm \rho}_3,$ respectively. The p.d.f. of the tumbling time is again exponential for long times with a time scale increasing exponentially with $\mathit{Wi}$ (figure \ref{fig:pdftumbling}, right). The tumbling statistics, therefore, shows properties similar to those found in a purely two dimensional flow. For the same values of $\mathit{Wi},$ the mean tumbling time is however significantly shorter than in two dimensions. This fact is attributed to the increased dimensionality of the system, which makes it easier for the trumbbell to escape from the aligned or anti-aligned configuration. For the same reason, the asymptotic exponential behavior of $\tau_0$ is observed at larger $\mathit{Wi}$ compared to the two-dimensional case.

\section{Conclusion}
\label{conclusion}

The phenomenon of tumbling is commonly associated with shear flows. The simplest objects that perform end-over-end tumbling in a simple shear, viz. axisymmetric solid particles and the
dumbbell, exhibit a trivial orientational dynamics in an extensional flow and  do not tumble. We have shown that, in contrast, a rich dynamics is obtained by considering one of the most elementary semi-flexible objects, namely the trumbbell. The mere consideration of one bending mode indeed yields a random end-over-end tumbling motion with exponentially distributed tumbling times. While it is well known that several microscopic objects tumble in a shear flow, 
to the best of our knowledge tumbling had not been observed in an extensional flow before. Moreover, we find a fundamental 
qualitative difference between the tumbling motion in an extensional flow and the analogous motion in a shear flow.
In the latter case, the typical tumbling time decreases as a power law of $\mathit{Wi}$; in the former case, it increases exponentially as $\mathit{Wi}$ increases. This difference is a consequence of the fact that the configurations aligned with the flow are stable in an extensional flow, whereas they are unstable in a shear flow.

We have also shown that, in an extensional flow, a trumbbell reverses its orientation by folding and then extending again in the opposite direction. This dynamics is reminiscent of the buckling instability of a fiber near to a hyperbolic point \cite{LS14}. Nevertheless, the tumbling motion described here is triggered by Brownian fluctuations that bend the trumbbell, whereas the buckling instability of a fiber results from its internal dynamics and does not necessarily require Brownian fluctuations.

The simplicity of the trumbbell model has allowed us to study its tumbling motion in detail and to relate this phenomenon to the properties of the stable configurations of the trumbbell and to the structure of the corresponding basins of attraction.
In our study, we have focused on the most elementary version of the trumbbell model. In particular, we have 
disregarded hydrodynamic and excluded volume interactions between the beads.
If the size of the beads is sufficiently smaller than their mutual separations,
the inclusion of such interactions into the model would somewhat modify the
stable folded configuration but not the essential structure of the phase space of
the system.
Therefore, although a more accurate description of the trumbbell would require taking into
account hydrodynamic and excluded volume interactions \cite{DdlT88,LAEP88}, 
we do not expect these interactions to alter the essential properties of the tumbling dynamics,
as long as the size of the beads is small compared to their mutual separations.

\enlargethispage{20pt}

%
%
%
\acknowledgements
{This work was supported in part by the EU COST Action MP 1305 `Flowing Matter'. ELCMP acknowledges the support of EACEA through the Erasmus Mundus Mobility with Asia program.}
{The authors are grateful to A. Ali, F. Delarue, S. Musacchio and S. S. Ray for useful discussions.}

\appendix
\section{Equations of motion of the trumbbell}
\label{appA}
In this appendix, we derive an alternative formulation of the internal dynamics of the trumbell that is suitable for numerical simulations in three dimensions. In the reference frame of ${\bm x}_{\mathrm{CM}}$, the configuration of a trumbbell can be described by using the vectors ${\bm \rho}_1={\bm x}_1-{\bm x}_2$ and ${\bm \rho}_3={\bm x}_3-{\bm x}_2.$ The position vectors of the beads satisfy:
\begin{equation}
 m \ddot{\bm x}_\nu =-\zeta[\dot{\bm x}_\nu-\bm u(\bm x_\nu,t)]+ \bm{\mathfrak{t}}_\nu+ \bm{f}_\nu + \sqrt{D}\,\bm \eta_\nu(t), \qquad \nu=1,2,3, \\
 \label{eq:langevin}
\end{equation}
where $m$ is the mass of each bead, $D=2\zeta k_BT$ and $\bm \eta_\nu(t)$ are independent $d$-dimensional white noises. The restoring forces $\bm{f}_\nu$ take the form $\bm{f}_1=\mu_\chi \left(\hat{\bm \rho}_1 \cot \chi-\hat{\bm \rho}_3 \csc \chi \right), \bm{f}_3=\mu_\chi \left(\hat{\bm \rho}_3 \cot \chi -\hat{\bm \rho}_1 \csc \chi \right)$   and  $\bm{f}_2=-(\bm{f}_1+\bm{f}_3)$ with $\mu_\chi=\mu(\pi-\chi)/\ell$ \cite{DdlT88}. The tensions $\bm{\mathfrak{t}}_\nu$ keep the distances between the beads constant. The terms on the r.h.s. of \eqref{eq:langevin} describe Stokes' drag, the rigidity of the rods, the resistance of the particle to bend, and Brownian noise, respectively. Neglecting inertial effects yields:
\begin{equation}
\dot{\bm x}_\nu =\displaystyle \bm u(\bm x_\nu,t) + \frac{1}{\zeta} \left[ \bm{\mathfrak{t}}_\nu+ \bm{f}_\nu + \sqrt{D}\, \bm \eta_\nu(t) \right], \qquad \nu =1,2,3. 
\end{equation}
By using the linearity of the velocity field, we obtain:
\begin{equation}
\dot{\bm x}_{\mathrm{CM}}=\bm u(\bm x_{\mathrm{CM}},t)+\sqrt{\dfrac{D}{3}}\,\hat{\bm \eta}(t),
\end{equation}
where $\hat{\bm\eta}(t)$ is white noise, and
\begin{equation}
\begin{split}
\dot{\bm \rho}_1 &= \displaystyle \bm \rho_1 \bm \cdot{\bm \nabla} \bm u+ \frac{1}{\zeta}\left(2\bm{\mathfrak{t}}_1+\bm{\mathfrak{t}}_3\right)+\frac{1}{\zeta}\left(2\bm{f}_1+\bm{f}_3\right)+ \Gamma\tilde{\bm \eta}_1, \\
\dot{\bm \rho}_3 &= \displaystyle \bm \rho_3 \bm \cdot{\bm \nabla} \bm u+ \frac{1}{\zeta}\left(2\bm{\mathfrak{t}}_3+\bm{\mathfrak{t}}_1\right)+\frac{1}{\zeta}\left(2\bm{f}_3+\bm{f}_1\right)+ \Gamma\tilde{ \bm \eta}_3,
\end{split}
\label{eq:drho}
\end{equation}
where $\Gamma=\sqrt{2D}/\zeta$ and $\tilde{\bm \eta}_i=\left[\bm \eta_i(t) - \bm \eta_2(t)\right]/\sqrt{2}$ with $\langle \tilde{ \bm \eta}_i(t) \rangle=0$ and $\langle \tilde{\eta}_i^{\alpha}(t) \tilde{\eta}_i^{\beta}(t') \rangle=\delta^{\alpha\beta}\delta(t-t'), i=1,3.$ We impose the rigidity constraints $d|\bm \rho_i|^2/dt=0$ to obtain: 
\begin{eqnarray}
\bm\rho_i \circ \dot{\bm\rho}_i=0,
\label{eq:rigidity}
\end{eqnarray}
where the symbol `$\circ$' indicates that the dot products involving the noise terms are understood in the Stratonovich sense \cite{KP99}. Equation \eqref{eq:rigidity} may now be used to calculate the tensions:
\begin{equation}
\begin{split}
\bm{\mathfrak{t}}_1 &=\displaystyle -\zeta c_\chi \,[2\sigma_1-\sigma_3 \cos \chi- \tilde{\mu}_\chi  +2\Gamma(\hat{\bm\rho}_1 \circ \tilde{\bm \eta}_1)-\Gamma \cos \chi (\hat{\bm\rho}_3 \circ \tilde{\bm \eta}_3) ]\hat{\bm\rho}_1, \\
\bm{\mathfrak{t}}_2 &= -(\bm{\mathfrak{t}}_1+\bm{\mathfrak{t}}_3), \\
\bm{\mathfrak{t}}_3 &=\displaystyle -\zeta c_\chi \,[2\sigma_3-\sigma_1 \cos \chi- \tilde{\mu}_\chi  +2\Gamma(\hat{\bm\rho}_3 \circ \tilde{\bm \eta}_3)-\Gamma \cos \chi (\hat{\bm\rho}_1 \circ \tilde{\bm \eta}_1) ]\hat{\bm\rho}_3,
\end{split}
\label{eq:tensions}
\end{equation}
where $\sigma_i =\sum_{\alpha,\beta}\ell\hat{\rho}_i^\alpha\partial^\alpha u^\beta \hat{\rho}_i^\beta$ with $\hat{\bm \rho}_i={\bm \rho}_i/|{\bm \rho}_i|, c_\chi=(4-\cos^2 \chi)^{-1},$ and $  \tilde{\mu}_\chi=\zeta^{-1}\mu_\chi\sin \chi(2- \cos \chi).$ By substituting \eqref{eq:tensions} into \eqref{eq:drho}, we obtain the system:
\begin{equation}
\begin{split}
\dot{\bm\rho}_1 &= \bm A_1 + \mathcal B_{11} \circ\tilde{ \bm \eta}_1(t)+ \mathcal B_{13} \circ\tilde{ \bm \eta}_3(t), \\
\dot{\bm\rho}_3 &= \bm A_3 + \mathcal B_{31} \circ\tilde{ \bm \eta}_1(t)+ \mathcal B_{33} \circ\tilde{ \bm \eta}_3(t).
\end{split}
\label{eq:stratform}
\end{equation} 
The explicit forms of the vectors $\bm A_1,\bm A_3$ are:
\begin{eqnarray}
\bm A_1 &=&\displaystyle \bm \rho_1 \bm \cdot {\bm \nabla}\bm u  -2c_\chi \left[2\sigma_1-\sigma_3 \cos \chi- \tilde{\mu}_\chi \right]\hat{\bm \rho}_1-c_\chi\left[2\sigma_3-\sigma_1 \cos \chi- \tilde{\mu}_\chi \right]\hat{\bm \rho}_3 +\frac{1}{\zeta}\left(2\bm{f}_1+\bm{ f}_3 \right), \nonumber \\ 
\bm A_3 &=&\displaystyle \bm \rho_3 \bm \cdot {\bm \nabla}\bm u - 2c_\chi \left[2\sigma_3-\sigma_1 \cos \chi- \tilde{\mu}_\chi \right]\hat{\bm \rho}_3 -c_\chi \left[2\sigma_1-\sigma_3 \cos \chi- \tilde{\mu}_\chi \right]\hat{\bm \rho}_1 
+\frac{1}{\zeta}\left(2\bm{f}_3+\bm{ f}_1 \right), \nonumber 
\end{eqnarray} 
and the matrix-valued coefficients of the noises are
\begin{eqnarray}
\mathcal B_{11} &=&\displaystyle \Gamma \left[\mathcal{I}+c_\chi\left(\hat{\bm \rho}_3 \cos \chi - 4\hat{\bm \rho}_1 \right)  \otimes\hat{\bm \rho}_1 \right], \nonumber \\
\mathcal B_{13} &=&\displaystyle 2\Gamma c_\chi \left(\hat{\bm \rho}_1 \cos \chi- \hat{\bm \rho}_3\right)  \otimes\hat{\bm \rho}_3, \nonumber \\
\mathcal B_{31} &=&\displaystyle 2\Gamma c_\chi \left(\hat{\bm \rho}_3 \cos \chi- \hat{\bm \rho}_1\right)  \otimes\hat{\bm \rho}_1, \nonumber \\
\mathcal B_{33} &=&\displaystyle \Gamma \left[\mathcal{I}+c_\chi\left(\hat{\bm \rho}_1 \cos \chi - 4\hat{\bm \rho}_3 \right) \otimes \hat{\bm \rho}_3 \right], \nonumber 
\end{eqnarray}  
where $\mathcal{I}$ is the identity matrix and $\otimes$ denotes the tensor product. Equations \eqref{eq:stratform} can be shown to be statistically equivalent to \eqref{eq:diffusion}. The angle $\chi$ can be obtained from ${\bm \rho}_1 \bm \cdot {\bm \rho}_3=\ell^2 \cos \chi.$ The above formulation holds both for $d=2$ and for $d=3.$



\end{document}